# Combination of OFDM and CDMA for high data rate UWB

Combinaison des techniques OFDM et CDMA pour l'UWB haut débit

Emeric Guéguen[*], Nadia Madaoui, Jean-François Hélard, Matthieu Crussière

Institute of Electronics and Telecommunications of Rennes (IETR)
INSA, 20 Avenue des Buttes de Coësmes, 35043 Rennes cedex, France

**Abstract**

For Wireless Personal Area Network (WPAN) systems, resource allocation between several users within a piconet and the coexistence of several piconets are very important points to take into consideration for the optimization of high data rate Ultra Wide Band (UWB) systems. To improve the performance of the Multi Band OFDM (Orthogonal Frequency Division Multiplex) solution proposed by the Multi Band OFDM Alliance (MBOA), the addition of a spreading component in the frequency domain is a good solution since it makes the resource allocation easier and also offers a better robustness against the channel frequency selectivity and narrowband interferences. The Spread Spectrum - Multi Carrier - Multiple Access (SS-MC-MA) system proposed in this paper offers not only the advantages of Multi Carrier - Coded Division Multiple Access (MC-CDMA) brought by frequency spreading but also a more effective dynamic resource allocation in a multi-user and multi-piconet context. These improvements are obtained without increasing the complexity of the radio-frequency part compared to the classical MBOA solution.

**Résumé**

Pour les systèmes WPAN (Wireless Personal Area Networks), la gestion des ressources entre plusieurs utilisateurs d'une même picocellule ainsi que la co-existence de plusieurs picocellules sont des points importants à prendre en compte lors de l'optimisation d'un système Ultra Large Bande (ULB) haut débit. Afin d'améliorer les performances de la solution Multi Band OFDM (Orthogonal Frequency Division Multiplex) proposée par l'alliance MBOA (Multi Band OFDM Alliance), l'ajout d'une composante d'étalement selon l'axe fréquentiel s'avère une bonne solution pour faciliter la gestion des ressources, qui offre en outre une meilleure robustesse vis-à-vis de la sélectivité en fréquence du canal et des interférences à bande étroite. Le système SS-MC-MA (Spread Spectrum - Multi Carrier - Multiple Access) que nous proposons, bénéficie non seulement des avantages du MC-CDMA (Multi Carrier - Coded Division Multiple Access) apportés par l'étalement fréquentiel mais permet également une allocation dynamique des ressources plus efficace dans un contexte multi-utilisateurs et multi-picocellules. Ces améliorations peuvent être obtenues, sans augmenter la complexité du segment radio-fréquence par rapport à la solution MBOA.

*Mots clés :* SS-MC-MA, ULB, MB-OFDM, Multi-utilisateurs, WPAN.

## 1. Introduction

Ultra Wide Band (UWB) radio systems are today acknowledged as high potential solutions for Wireless Personal Area Networks (WPAN). Many UWB system approaches are studied and proposed regarding different objectives in terms of service (localization or communication), distance ranges (short or middle) or data rates (high or low) leading to different waveforms. The novelty of these systems consists in the possibility of a non regulated access to the spectral resource, leading to a flexible use of the radio channel for an important number of applications. The standardization process, led by the Task Group 802.15.3a of the Institute of Electrical and Electronics Engineers (IEEE) to define a high data rate physical layer for these WPAN, has seen these three last years the emergence and the confrontation of many solutions designated as ultra wide band solutions. Particularly, the solution known as Multi Band OFDM (Orthogonal Frequency Division Multiplex), considered by the Multi Band OFDM Alliance (MBOA) consortium, is currently promoted by the main actors of the general public and components industries [1,2]. It actually presents some advantages but also some drawback. After a critical analysis of the MBOA solution, this paper studies the interest of the addition of a Coded Division Multiple Access (CDMA) component to a Multi Band OFDM signal. Particularly, it is demonstrated that the proposed solution based on the combination of OFDM and CDMA technologies offers, for the future WPAN, good performance and a great flexibility for the resource allocation between users of a same piconet.

---
[*] *Corresponding author.*
  *E-mail address : emeric.gueguen@insa-rennes.fr (E. Guéguen)*

**2. Diffusion constraints and MBOA solution**

In 2002, the Federal Communications Commission (FCC) regulated UWB systems by imposing the spectral mask, illustrated in Fig. 1, to limit transmission power. To be considered as UWB, a signal must have a minimum bandwidth of 500 MHz or a bandwidth to central frequency ratio above 0,2. The power spectral density (PSD) must not exceed -41,3 dBm/MHz. The UWB channel, running from 3,1 to 10,6 GHz, is frequency selective and considered as almost invariant in time. Proposed UWB systems must not disturb existing narrowband systems, like for example Wireless Local Area Network 802.11a standard at 5 GHz.

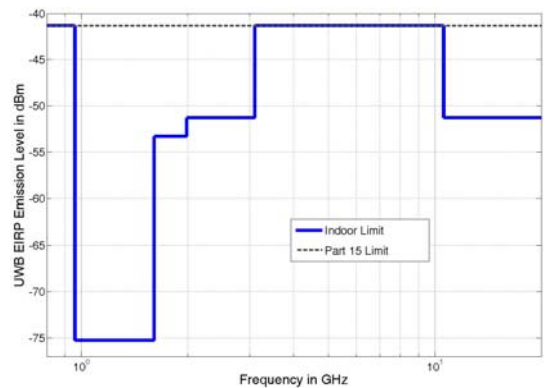

Fig. 1. Indoor PSD mask of the FCC
Fig. 1. Masque de DSP indoor de la FCC

Nowadays, within the IEEE 802.15.3a standardization authorities, two main approaches have been proposed for the high data rate UWB: a pulse radio solution using DS-CDMA (Direct Sequence-CMDA) ternary codes and the multi-carrier multi-band solution [1]. The latter, supported by the MBOA consortium, proposes to divide the available band into 14 sub-bands of 528 MHz each, as illustrated in Fig. 2.

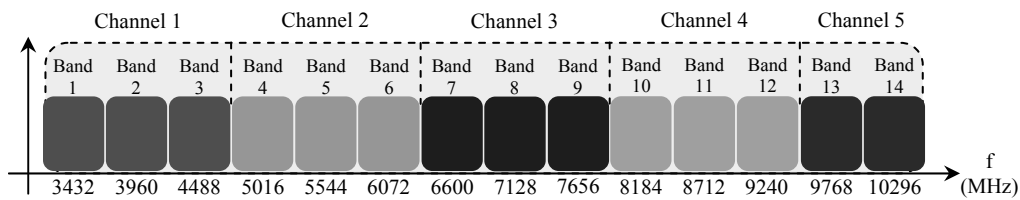

Fig. 2. Channels distribution for MBOA solution
Fig. 2. Organisation des modes de la solution MBOA

Each of these sub-bands allows the transmission of an OFDM signal, obtained from a 128 points Inverse Fast Fourier Transform (IFFT). In the first step, studies are focused on the first mode which clusters the first three sub-bands from 3,1 to 4,8 GHz. The multi-user management within a piconet is based on Time Division Multiple Access (TDMA) by using a Time-Frequency Code (TFC). At a given moment, each user occupies one of the three sub-bands of mode 1 [2]. The signal, sampled during the analog to digital conversion, has a limited bandwidth of 500 MHz, leading to low cost components. However, the TFC application introduces frequency hopping from one sub-band to another at the end of each OFDM symbol, thus allowing each user to benefit from frequency diversity owing to the width of three sub-bands in mode 1. In addition, considering that each user occupies a given sub-band only one third of time, it is possible to optimize the transmitted power while respecting the PSD mask imposed by the FCC. Lastly, it is also advised to plan the cohabitation of 4 piconets in a same environment.

**Main parameters of the MBOA solution**

Transmitted data rates in each sub-band depend firstly, on the coding rate, as the applied modulation to the different subcarriers of the OFDM multiplex is a quadrature phase-shift keying (QPSK). These different modes corresponding to data rates from 53,3 to 480 Mbit/s are enumerated in Tab. 1.

| Data Rate (Mbit/s) | Modulation | Coding Rate (R) | Conjugate Symmetric Input to IFFT | Time Spreading Factor (TSF) | Coded bits per OFDM symbol ($N_{CBPS}$) |
|---|---|---|---|---|---|
| 53,3 | QPSK | 1/3 | Yes | 2 | 100 |
| 80 | QPSK | 1/2 | Yes | 2 | 100 |
| 110 | QPSK | 11/32 | No | 2 | 200 |
| 160 | QPSK | 1/2 | No | 2 | 200 |
| 200 | QPSK | 5/8 | No | 2 | 200 |
| 320 | QPSK | 1/2 | No | 1 (No spreading) | 200 |
| 400 | QPSK | 5/8 | No | 1 (No spreading) | 200 |
| 480 | QPSK | 3/4 | No | 1 (No spreading) | 200 |

Tab. 1. Data rates in different modes - MBOA solution
Tab. 1. Débits transmis dans les différents modes - solution MBOA

The channel code is a 64 states convolutional code with a rate of 1/3 (with generator polynomials $g_0 = 133_8$, $g_1 = 165_8$ and $g_2 = 171_8$) which is punctured to obtain higher rates. For certain modes, each complex symbol and its conjugate symmetric is transmitted into the same OFDM symbol by one subcarrier and its "mirror" subcarrier respectively. This way, the frequency diversity is exploited into each sub-band at the cost of a division by 2 of the useful transmitted data rate. Moreover, for modes corresponding to data rates from 53,3 Mbit/s to 200 Mbit/s, a time spreading of 2 is applied. It consists in the transmission of the same information during 2 consecutive OFDM symbols in order to take advantage of a better frequency diversity, due to the joint application of the TFC. Finally, an interleaving on binary data generated by the encoder is applied in three steps: the symbol block interleaver permutes binary elements within the span of 6 OFDM symbols, the tone block interleaver permutes binary elements transmitted by the subcarriers of an OFDM symbol and the intra-symbol cyclic shifts which consists of different cyclic shifts of symbol blocks within the span of 6 symbols.

The main OFDM parameters of the MBOA solution are summarized in Tab. 2.

| Parameter | Value |
|---|---|
| IFFT/FFT size | 128 |
| Sampling frequency | 528 MHz |
| Transmission bandwidth | 507,37 MHz |
| Number of data subcarriers | 100 |
| Number of pilot subcarriers | 12 |
| Number of guard subcarriers | 10 |
| $N_{ST}$ : Total number of used subcarriers | 122 |
| $\Delta_F$ : Sub-carrier frequency spacing | 4,125 MHz (=528MHz/128) |
| $T_{FFT}$ : IFFT/FFT period | 242,42 ns ($1/\Delta_F$) |
| $T_{CP}$ : Zero Paddind prefix duration | 60,61 ns |
| $T_{GI}$ : Zero Paddind guard interval duration | 9,47 ns |
| $T_{SYM}$ : Symbol interval | 312,5 ns ($T_{FFT} + T_{CP} + T_{GI}$) |

Tab. 2. Main OFDM parameters - MBOA solution
Tab. 2. Principaux paramètres OFDM - solution MBOA

The IFFT size is 128 and the total number of used subcarriers is 122. The useful duration of each OFDM symbol is 242 ns, leading to subcarrier frequency spacing of $\Delta_F = 4,125$ MHz. A zero padding (ZP) guard interval of 60,61 ns duration is added at the end of each OFDM symbol to cope with inter-symbol interference.

The only difference of ZP with the traditional cyclic prefix (CP) is that the CP is replaced by D trailing zeros. This operation, performed on OFDM symbols at the IFFT input, is presented by equation (1):

$$\widetilde{s}_{ZP}(i) = F_{ZP} s_M(i) \qquad (1)$$

where $F_{ZP} = \begin{bmatrix} F_M^H \\ 0_{D \times M} \end{bmatrix}$, $F_M$ is the $M \times M$ FFT matrix and $(.)^H$ denotes conjugate transposition, so $F_M^H$ is the IFFT matrix. $s_M(i)$ is the i$^{th}$ $M \times 1$ OFDM symbol at the IFFT input and $M$ is the OFDM symbol length at the IFFT input, in this case $M = 128$.

The input receiver signal is given by:

$$\widetilde{x}_{ZP}(i) = HF_{ZP}s_M(i) + H_{IBI}F_{ZP}s_M(i-1) + \widetilde{n}_P(i) \quad (2)$$

where $H$ is the $P \times P$ ($P = M + D$) lower triangular Toeplitz matrix with the first column $[h_0 \ldots h_L 0 \cdots 0]^T$, $H_{IBI}$ is the $P \times P$ upper triangular Toeplitz matrix with the first row $[0 \cdots 0\, h_L \cdots h_1]$ which symbolizes inter-block interference (IBI). The frequency-selective propagation is modelled as a FIR filter with channel impulse response column vector $h = [h_L \cdots h_0]^T$ where $L$ is the channel order ($L \leq D \leq M$). $\widetilde{n}_P(i)$ is the $P \times 1$ additive white Gaussian noise (AWGN) vector.

The all-zero $D \times M$ matrix $0_{D \times M}$ eliminates the IBI, since $H_{IBI}F_{ZP} = 0_{P \times M}$. Let $H = [H_0, H_{ZP}]$ denotes a partition of the $P \times P$ matrix $H$ where $H_0$ is the matrix made of the $M$ first columns of $H$ and $H_{ZP}$ the matrix made of the $D$ last columns. The received $P \times 1$ vector then becomes

$$\widetilde{x}_{ZP}(i) = HF_{ZP}s_M(i) + \widetilde{n}_P(i) = H_0 F_M^H s_M(i) + \widetilde{n}_P(i) \quad (3)$$

We can split $\widetilde{x}_{ZP}(i)$ in (3) into its upper $M \times 1$ part $\widetilde{x}_u(i) = H_u \widetilde{s}_M(i)$ and its lower $D \times 1$ part $\widetilde{x}_l(i) = H_l \widetilde{s}_M(i)$ where $H_u$ (respectively $H_l$) denotes the corresponding $M \times M$ (resp. $D \times M$) partition of $H_0$. So we can form

$$\begin{aligned}\widetilde{x}_M(i) &= \widetilde{x}_u(i) + \begin{bmatrix} \widetilde{x}_l(i) \\ 0_{(M-L) \times 1} \end{bmatrix} \\ &= \left( H_u + \begin{bmatrix} H_l \\ 0_{(M-L) \times M} \end{bmatrix} \right) \widetilde{s}_M(i) \\ &= C_M(h) \widetilde{s}_M(i) \end{aligned} \quad (4)$$

where $\widetilde{s}_M(i)$ is the i[th] OFDM symbol at the IFFT output and $C_M(h)$ is $M \times M$ circulant matrix with the first row $C_M(h) = Circ_M(h_0\, 0 \cdots 0\, h_L \cdots h_1)$. The used of ZP prefix requires to make the operation presented by the equation (4) and called Overlap and Add (OLA) [3] before the FFT demodulation in order to restore the orthogonality between subcarriers of the multiplex. Also, the obtained result, $\widetilde{x}_M(i)$, is the temporal signal at the FFT input i.e. after correct windowing and OLA operation. The other parts of the receiver are strictly identical to those used with the traditional CP.

ZP allows obtaining a spectrum with fewer ripples in the useful band than with a traditional CP. Thus the signal can take the exact shape of the PSD mask. The signal shape is optimized but at the cost of an additional processing in the receiver.
Moreover, an additional guard interval, also of ZP type, of 9,47 ns is added for the commutation from one sub-band to another. The complete solution of the MBOA solution is described in details in [1], and an accurate analysis of its performances is given in [4] in the case of a perfect and real channel estimation.

To summarize, the MBOA solution offers some advantages for high data rate UWB applications, such as the signal robustness against the channel selectivity and the efficient exploitation of the signal energy received within the prefix duration. The main argument of multi-carrier modulation in general is often quoted in favour of the MBOA solution, when one compares it with the competitive DS-CDMA solution. Indeed, the latter can make use of all the received energy with difficulty, the RAKE finger number being compulsorily limited for complexity reasons. However, the freedom degrees of the MBOA solution are relatively limited in a multi-user and multi-piconet context. Particularly, when only the three first sub-bands of the first mode are considered, conflicts appear immediately with the fourth user within a piconet, whereas scenarios going up to 6 simultaneous users have classically to be considered.

**3. Why adding a CDMA component to the MBOA solution?**

Starting from the MBOA solution, some studies have already proposed to add a CDMA component in order to improve the system robustness or the resource sharing between several users. Indeed, this spreading component allows particularly to organize the access of several users to a common resource. Taking into account the UWB channel characteristics, frequency selectivity and slow time variation in indoor environment, spreading is generally performed along the frequency axis, leading to MC-CDMA signals [5,6,7,8]. The symbols of all users are then transmitted by all the subcarriers, the spreading codes length being lower or equal to the subcarriers number of the OFDM multiplex (Fig. 3). By another way, [9] compares an OFDM system with an MC-CDMA system for UWB applications at 60 GHz. Generally, these different contributions show that, compared to the "traditional" MBOA solution, beyond a greater facility in the resource sharing in multi-user case, an MC-CDMA system also presents a better robustness against the channel frequency selectivity. Moreover, a spreading component in the frequency domain improves the UWB signal robustness against narrowband interferences. This last point is fundamental for uncontrolled access to the spectral resource considering a flexible use of the radio channel for a great number of different applications. However, in a particular case [5], authors suggest to use an MC-CDMA signal with a bandwidth $B_W$ = 1,58 GHz, equivalent to 3 sub-bands of the MBOA signal, which leads to highly increase the sampling frequency of the analog-to-digital conversion.

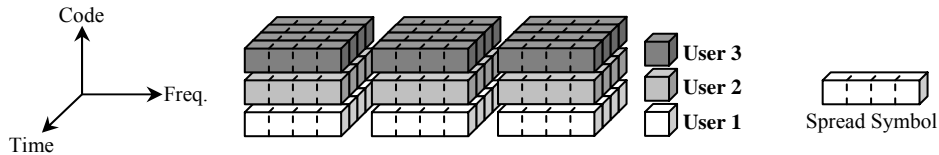

Fig. 3. MC-CDMA principle
Fig. 3. Illustration du principe du MC-CDMA

## 4. A new waveform for multi-band UWB: the SS-MC-MA

By applying a spreading code and a multi-access component, we propose in this article an SS-MC-MA (Spread Spectrum – Multi Carrier – Multiple Access) waveform [10], which is new for UWB applications and offers better performance and more flexibility in the resource management. While considering the WPAN context and UWB channel characteristics, we choose a frequency domain spreading solution. In addition, particularly for technological constraints due to the analog-to-digital conversion, the bandwidth of the transmitted signal at a given moment will be limited to 528 MHz and not 1,584 GHz (equivalent to 3 sub-bands of the first mode) as proposed in [5]. By using a frequency hopping technique over the 3 sub-bands, as proposed by the MBOA solution, it is possible to benefit from the frequency independence due to a bandwidth equal to 1,584 GHz.

*4.1. SS-MC-MA principle*

With MC-CDMA, resource sharing is realized by the assignment to each user of one or more codes, which are transmitted all over the available bandwidth. Thus, all subcarriers of the whole allocated spectrum transmit the symbols of all users differentiated by their individual code. Then, MC-CDMA can be view as a mono-block system (Fig. 3).

In a multi-block system, spectrum is divided into "blocks" of many subcarriers. Among the possible combinations, the SS-MC-MA solution, illustrated in Fig. 4, consists in assigning to each user a specific block of subcarriers according to a Frequency Division Multiple Access (FDMA) scheme. Code dimension can then be exploited for an adaptive resource optimization and sharing (modulation type, data rate…). Spreading in the frequency domain leads to diversity gain and, as it is the case for a MC-CDMA signal, improves the UWB signal robustness against narrowband interferers. With a SS-MC-MA signal, symbols are transmitted simultaneously on a specific subset of subcarriers by the same user and undergo the same distortions. Self-interference (SI), which then replaces the multiple access interference (MAI) obtained with MC-CDMA signals, can be easily compensated by mono-user detection with only one complex coefficient per subcarrier.

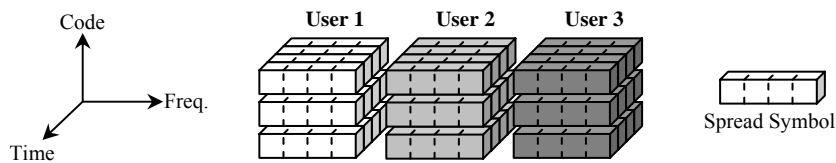

Fig. 4. SS-MC-MA system principle illustration
Fig. 4. Illustration du principe du système SS-MC-MA

*4.2. SS-MC-MA advantages compared with MBOA solution and MC-CDMA*

Let us consider the case of the MBOA standard in mode 1 (Fig. 2).
- Case of three or less users
  In this case, the SS-MC-MA system allows the allocation of a 528 MHz sub-band for each user. This system offers the same performance and advantages as MC-CDMA, while bringing an additional freedom degree compared with the MBOA solution for the dynamic resource allocation, via the allocation of a given number of spreading codes. Another advantage, compared in this case with an MC-CDMA system, is the simpler channel estimation in reception. Indeed, a given subcarrier is distorted by only one channel, the one of the user associated with this subcarrier. With an MC-CDMA system, each subcarrier is corrupted by the different channels of different users, which increases in a considerable way the channel estimation complexity. In that case, each user has to estimate the response of many channels all over the total available bandwidth.
- Case of more than three users
  In the MBOA solution, conflicts appear from 4 users and could cause information losses. In the SS-MC-MA case, code dimension could be exploited to share a same 528 MHz sub-band between 2 or even 3 users if necessary. In that case, the generated signal within a given block corresponds to a MC-CDMA signal, but with a limited number of users per block (2 or even 3).

More generally, in a multi-piconet context, flexibility brought by the resource sharing, by modifying the number of spreading codes assigned to a given user in a given piconet, allows the SS-MC-MA system to offer a more efficient dynamic resource allocation than the MBOA solution.

**5. Proposed system description**

*5.1. Studied system*

The proposed system is very similar to the MBOA one. Fig. 5 introduces the MBOA transmission chain in continuous lines and, in dashed lines, the added functions to obtain an SS-MC-MA signal. These functions are mainly the Hadamard Transform ("Fast Hadamard Transform": FHT) at the transmitter and the inverse transform (Inverse FHT) at the receiver. In addition, Zero Forcing (ZF) or Minimum Mean Square Error (MMSE) single user detection is applied.

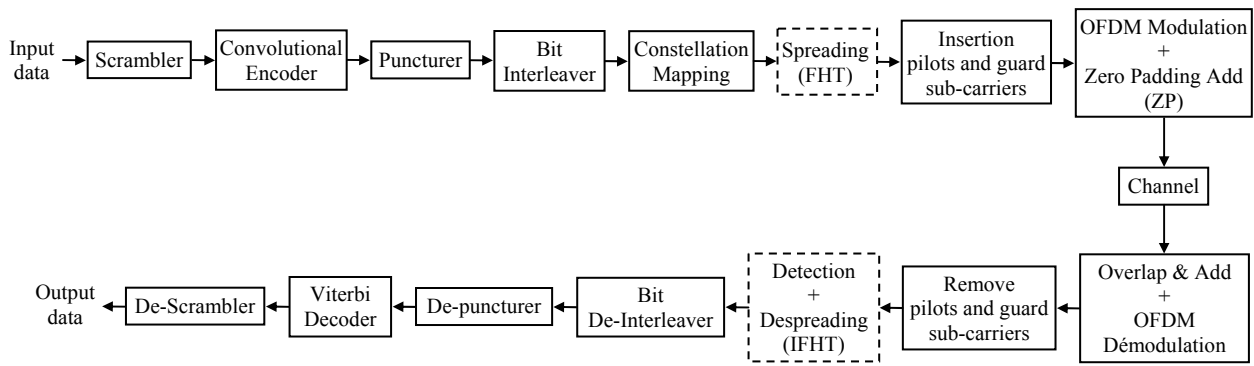

Fig. 5. MBOA transmission chain (SS-MC-MA in broken lines)
Fig. 5. Chaîne de transmission MBOA (SS-MC-MA en traits pointillés)

The spreading length $L_c$ is chosen equal to 16 and the number of data subcarriers is reduced from 100 to 6x16 = 96 for each OFDM symbol. That means that 4 more guard subcarriers are added. Other functions stay unchanged, particularly the OLA operation before the FFT demodulation to restore the orthogonality between subcarriers.

*5.2. Expression of the signals*

In the MBOA solution case, the signal generated at the output of the inverse FFT is equal to:

$$S_{OFDM}(t) = \sum_{i=-\infty}^{+\infty} \sum_{n=-N_{ST}/2}^{n=-N_{ST}/2} X_n(i) p_c(t - iT_{CP}) e^{j2\pi n \Delta_F (t - iT_{CP})} \quad (5)$$

where $\Delta_F$, $N_{ST}$ and $T_{SYM}$ represent subcarriers spacing, the number of total used subcarriers and the spacing between two consecutives OFDM symbols respectively. $X_n(i)$ is a complex symbol, belonging to a QPSK constellation and is transmitted by subcarrier $n$ during the $i^{th}$ OFDM symbol. It represents a data, a pilot or a reference symbol. $p_c(t)$ is a rectangular window defined by:

$$p_c(t) = \begin{cases} 1, & 0 \leq t \leq T_{FFT} + T_{CP} \\ 0, & T_{FFT} + T_{CP} \leq t \leq T_{FFT} + T_{CP} + T_{GI} \end{cases} \quad (6)$$

In the SS-MC-MA case, complex symbols are converted into $P$ parallel symbols $D_l(i)$ (with $P \leq L_c$), which are transmitted by the same $L_c$ subcarriers. $P$, representing the load, is for example equal to $L_c$ in the full load case and equal to $L_c/2$ in the half load. $C_l = [c_{l,1} \ldots c_{l,m} \ldots c_{l,Lc}]$ is the $l^{th}$ Walsh-Hadamard orthogonal spreading code. In this case, the waveform is the same as previously, but the complex symbol $X_m(i)$ which is transmitted by the $m^{th}$ subcarrier ($m$ varying from 1 to $L_c = 16$, with $m = n$ modulo(16)) of a block of $L_c = 16$ carriers bound by the same spreading codes of length $L_c$ can be express by :

$$X_m(i) = \sum_{l=1}^{l=P} D_l(i) c_{l,m} \quad (7)$$

where $D_l(i)$ represents the $P$ complex symbols, belonging to a QPSK constellation and which are transmitted by the block of $L_c$ subcarriers considered during the OFDM symbol $i$. In reception, as in the classical OFDM system case, mono-user detection is simply realized at the output of the FFT by one complex multiplication per subcarrier. ZF and MMSE detection techniques are considered leading to coefficients respectively given by:

$$g_{n,i} = \frac{1}{h_{n,i}} \quad \text{Zero Forcing}$$

$$g_{n,i} = \frac{h_{n,i}^*}{|h_{n,i}|^2 + \frac{1}{\gamma_{n,i}}} \quad \text{MMSE} \quad (8)$$

where $h_{n,i}$ and $\gamma_{n,i}$ represent the complex channel response and the signal to noise ratio for the subcarrier $n$ of the symbol $i$ respectively.

*5.3. UWB channel modelisation*

Wideband propagation channels which are used for UWB systems PHY layer evaluation, result from Saleh-Valenzuela model for indoor applications [11]. This ray based model takes into account clusters phenomena highlighted during channels measurements. The multipath channel impulse response for the $k^{th}$ user is given by:

$$h_k(t) = \sum_{m=0}^{M_k} \sum_{p=0}^{P_k} \alpha_k(m,p) \delta(t - T_k(m) - \tau_k(m,p)) \quad (9)$$

where $T_k(m)$ is the cluster $m$ delay, $\alpha_k(m,p)$ and $\tau_k(m,p)$ are the gain and the delay of the path $p$ of the cluster $m$ respectively. The mean excess delay $\tau_m$, and the root mean square delay spread, $\tau_{rms}$, are given in Tab. 3 for the 4 channel models CMi (i = {1,…,4}).

| Characteristics | CM1 | CM2 | CM3 | CM4 |
|---|---|---|---|---|
| Mean excess delay (ns) : $\tau_m$ | 5,05 | 10,38 | 14,18 | |
| Root mean square delay spread: $\tau_{rms}$ | 5,28 | 8,03 | 14,28 | 25 |
| Distance (m) | < 4 | < 4 | 4 - 10 | 10 |
| LOS/NLOS | LOS | NLOS | NLOS | NLOS |

Tab. 3. Characteristics of wideband channels CMi
Tab. 3. Caractéristiques des canaux large bande CMi

In the Line Of Sight (LOS) configuration, transmitter and receiver antennas are in direct visibility, contrary to the Non Line Of Sight (NLOS) configuration. Under the assumption that the prefix duration is higher than the channel impulse response spreading and that the subcarriers number $N_{ST}$ is sufficiently large to guarantee the frequency non-selectivity of the channel for each one of these subcarriers, UWB channel can be modelled in the form of $N_{ST}$ sub-channels. In this case, the frequency response for the subcarrier $n$ for the $k^{th}$ user is given by:

$$H_k(n) = \sum_{m=0}^{M_k}\sum_{p=0}^{P_k} \alpha_k(m,p) e^{-j2\pi n \Delta_F (T_k(m)+\tau_k(m,p))} \quad (10)$$

where $\Delta_F$ is the subcarrier spacing.

The channel is thus modelled in the frequency domain, and for each sub-band, it is normalized in mean energy for each realization. 100 different realizations are used for each sub-band, one realization being applied along the frame duration. Typical realizations of the frequency responses of CM1 and CM4 models are represented in Fig. 6. As expected, we can note that the frequency selectivity is higher with CM4 than with CM1.

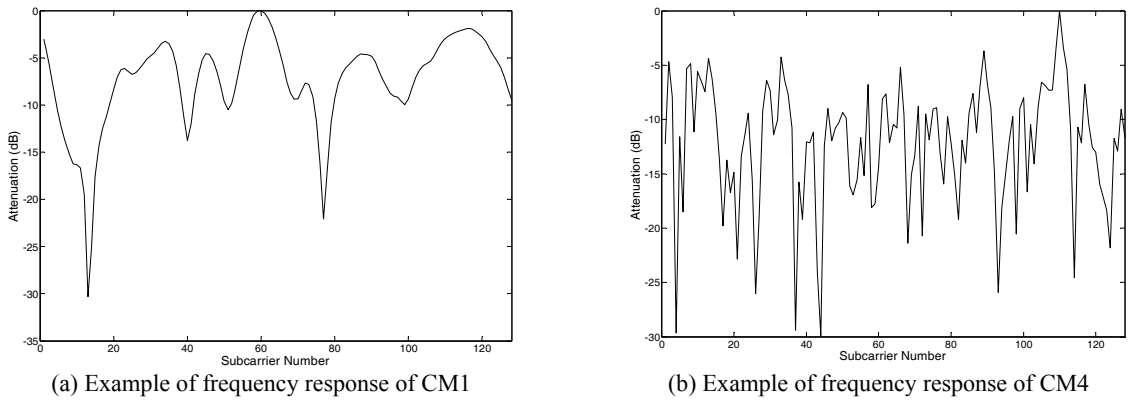

(a) Example of frequency response of CM1    (b) Example of frequency response of CM4

Fig. 6. Realizations of UWB models channels CM1 and CM4
Fig. 6. Réalisations des models de canaux UWB CM1 et CM4

## 6. Systems performance

Firstly, the performance of the MBOA has been estimated for the different UWB channels. Fig. 7a and 7b exhibit the results obtained with both channels CM1 and CM4 in the ideal case of perfect channel estimation and for rates ranged from 480 to 53,3 Mbits/s. These results are given versus $E_b/N_0$ which is the ratio between energy per useful bit and the noise monolateral spectral density. Note that ZP has been used in simulations and that have thus been considered to compute $E_b/N_0$. The performance of the MBOA system for a gaussian channel with a rate of 320 Mbit/s is also mentioned as a reference.

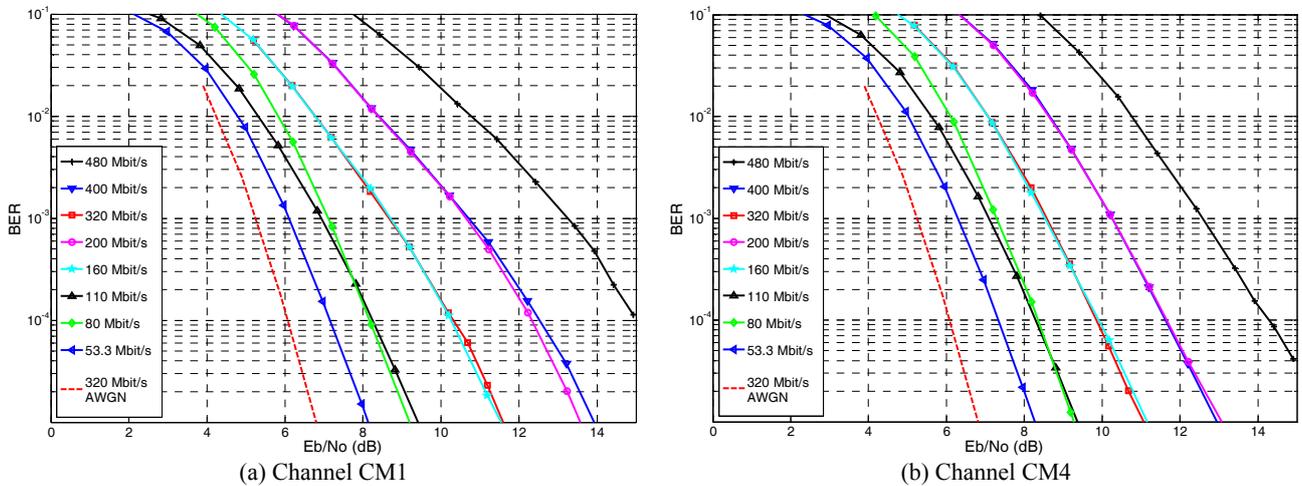

(a) Channel CM1    (b) Channel CM4

Fig. 7. Performance of the MBOA system
Fig. 7. Performances du système MBOA

For each mode, the performances with the two channels are very close. Results are slightly better with CM4 channel, because in that case the received signal benefits from a better frequency diversity. However the frequency domain representation of the channel does not take into account the scarce paths effects that could run over the guard interval and consequently could bring inter-carrier interference. Results versus the $E_b/N_0$ ratio for the rates of 320 Mbit/s and 160 Mbit/s are almost the same. For the rate of 160 Mbit/s, the spreading over two consecutive symbols does not bring any diversity, as the channel is the same for the two considered symbols.

With the SS-MC-MA system, the number of data subcarriers is reduced from 100 to 96 in order to obtain a multiple of the spreading length equal to 16. The total number $N_{ST}$ of used subcarriers is henceforth 118 and the transmission bandwidth becomes equal to 490,87 MHz instead of 507,37 MHz. Consequently, with a channel code rate equal to ½, the resulting useful data rate is 307 Mbit/s for full load SS-MC-MA system, with $P = L_c = 16$. The SS-MC-MA waveform brings an additional freedom degree for resource allocation. Via the attribution of a given number $P$ of spreading codes, it is very easy to obtain the wanted rate, as shown in Tab. 3 which gives the different data rates. For example, the 153 Mbit/s rate is obtained with a channel code rate equal to ½ and a half load system with $P = 8$, and the 76 Mbit/s rate corresponds to $P = 4$. Furthermore, neither conjugate symmetric function nor time spreading are applied.

| Data Rate (Mbit/s) | Modulation | Coding Rate (R) | Load $P$ | Coded bits per OFDM symbol ($N_{CBPS}$) |
|---|---|---|---|---|
| 57 | QPSK | 1/2 | 3 | 36 |
| 76 | QPSK | 1/2 | 4 | 48 |
| 153 | QPSK | 1/2 | 8 | 96 |
| 307 | QPSK | 1/2 | 16 | 192 |
| 409 | QPSK | 2/3 | 16 | 192 |
| 460 | QPSK | 3/4 | 16 | 192 |

Tab. 3. Data rates in different modes – SS-MC-MA solution
Tab. 3. Débits transmis dans les différents modes - solution SS-MC-MA

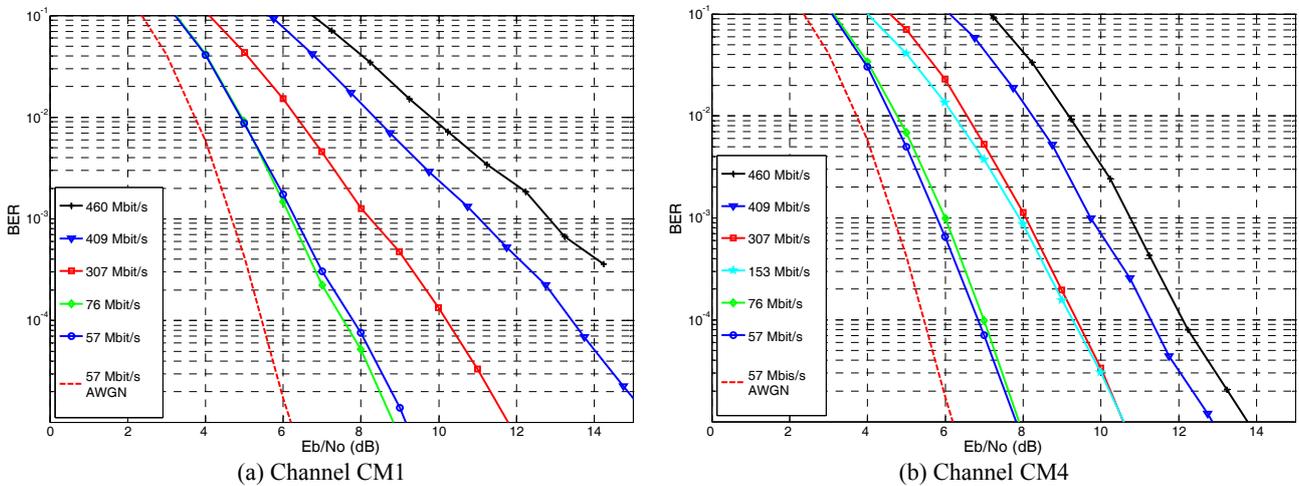

(a) Channel CM1  (b) Channel CM4

Fig. 8. Performance of the SS-MC-MA system
Fig. 8. Performances du système SS-MC-MA

|  | MBOA | | SS-MC-MA | |
|---|---|---|---|---|
|  | 80 Mbit/s | 320 Mbit/s | 76 Mbit/s | 307 Mbit/s |
| **CM1** | 8,2 dB | 10,3 dB | 7,6 dB | 10,2 dB |
| **CM4** | 8,4 dB | 9,85 dB | 7,0 dB | 9,4 dB |

Tab. 4. Required $E_b/N_0$ for BER = $10^{-4}$ versus the bit rate for the MBOA and SS-MC-MA systems
Tab. 4. Rapport $E_b/N_0$ nécessaire pour un TEB = $10^{-4}$ en fonction du débit pour les systèmes MBOA et SS-MC-MA

The performances of the SS-MC-MA scheme with MMSE single user detection are shown on Fig. 8a and 8b for the two channels CM1 and CM4. Note that the spreading codes have not necessarily been chosen optimally in simulations. As evident from the results obtained at 76 Mbit/s and 57 Mbit/s, the used codes are not favourable regarding the SI, which implies that the 76 Mbit/s scenario outperforms the 57 Mbit/s one. Moreover, a comparison of the two systems is presented in Tab. 4, which gives the required $E_b/N_0$ to obtain a BER equal to $10^{-4}$ for two different bit rates of the MBOA and SS-MC-MA systems for the channel CM1 and CM4. For example, the results obtained with the SS-MC-MA scheme at 76 Mbit/s ($P = 4$) are more attractive than those obtained with the MBOA system at 80 Mbit/s. These results show that the solution of the MBOA consortium, time spreading and conjugate symetric, to improve the performance of the low bit rate modes are not the most efficient. At higher bit rates, for example at 307 Mbit/s for SS-MC-MA and 320 Mbit/s for MBOA, this tendency turns out to be less perceptible and both systems perform very close to each other. It is eventually highlighted that the proposed SS-MC-MA solution can be worth of interest, namely for low bit rates.

**7. Conclusion**
To improve the performance of the MBOA solution and especially for a better resource allocation in a multi-user context, an adding spreading component in the frequency domain is a good solution which offers a better robustness to cope with the channel frequency selectivity and narrowband interferers. The proposed SS-MC-MA scheme has the advantages of MC-CDMA which are brought by frequency spreading, and also allows a more effective dynamic resource allocation in a multi-user and multi-piconet context. These improvements could be obtained without increasing the system complexity in comparison with the reference MBOA solution. Especially, the bandwidth of the received signal being still equal to 500 MHz, the radio-frequency part constraints are unchanged compared to the MBOA solution.

**Acknowledgements**
Authors would like to thank France Télécom R&D/RESA/BWA which supports this study within the contract 461 365 82.